\documentclass[aps,prl,reprint,superscriptaddress,floatfix]{revtex4-1}
\usepackage{braket}
\usepackage{amsmath}
\usepackage{graphicx}
\usepackage{placeins}
\usepackage{xcolor}
\graphicspath{{./figures/}}
\usepackage[]{hyperref}
\hypersetup{
    pdftitle={Fast State Transfer and Entanglement Renormalization Using Long-Range Interactions},
    pdfauthor={Zachary Eldredge},
    bookmarksnumbered=true,     
    bookmarksopen=true,         
    bookmarksopenlevel=1,      
    colorlinks=true,
    citecolor=blue,
    pdfstartview=Fit,           
    pdfpagemode=UseOutlines,    
    pdfpagelayout=TwoPageRight
}

\newcommand{\dd}[1]{\mathrm{d} #1}
\newcommand{\abs}[1]{\ensuremath{\left| #1 \right|}}

\begin{document}
\title{Fast State Transfer and Entanglement Renormalization Using Long-Range Interactions}
\date{\today}
\author{Zachary Eldredge}
\affiliation{Joint Quantum Institute and Joint Center for Quantum Information and Computer Science, NIST/University of Maryland, College Park, Maryland 20742, USA}
\author{Zhe-Xuan Gong}
\affiliation{Joint Quantum Institute and Joint Center for Quantum Information and Computer Science, NIST/University of Maryland, College Park, Maryland 20742, USA}
\author{Jeremy T. Young}
\affiliation{Joint Quantum Institute and Joint Center for Quantum Information and Computer Science, NIST/University of Maryland, College Park, Maryland 20742, USA}
\author{Ali Hamed Moosavian}
\affiliation{Joint Quantum Institute and Joint Center for Quantum Information and Computer Science, NIST/University of Maryland, College Park, Maryland 20742, USA}
\author{Michael Foss-Feig}
\affiliation{Joint Quantum Institute and Joint Center for Quantum Information and Computer Science, NIST/University of Maryland, College Park, Maryland 20742, USA}
\affiliation{United States Army Research Laboratory, Adelphi, Maryland 20783, USA}
\author{Alexey V. Gorshkov}
\affiliation{Joint Quantum Institute and Joint Center for Quantum Information and Computer Science, NIST/University of Maryland, College Park, Maryland 20742, USA}

\begin{abstract}
	In short-range interacting systems, the speed at which entanglement can be established between two separated points is limited by a constant Lieb-Robinson velocity. Long-range interacting systems are capable of faster entanglement generation, but the degree of the speed-up possible is an open question. In this paper, we present a protocol capable of transferring a quantum state across a distance $L$ in $d$ dimensions using long-range interactions with strength bounded by $1/r^\alpha$. If $\alpha < d$, the state transfer time is asymptotically independent of $L$; if $\alpha = d$, the time scales logarithmically with the distance $L$; if $d < \alpha < d+1$, transfer occurs in time proportional to $L^{\alpha - d}$; and if $\alpha \geq d + 1$, it occurs in time proportional to $L$. We then use this protocol to upper bound the time required to create a state specified by a MERA (multiscale entanglement renormalization ansatz) tensor network and show that if the linear size of the MERA state is $L$, then it can be created in time that scales with $L$ identically to state transfer up to logarithmic corrections. This protocol realizes an exponential speed-up in cases of $\alpha = d$, which could be useful in creating large entangled states for dipole-dipole ($1/r^3$) interactions in three dimensions.
\end{abstract}

\maketitle

Entanglement generation in a quantum system is limited, even in a non-relativistic setting, by the available interactions. In a lattice system with short-range interactions, Lieb and Robinson showed that there exists a linear light cone defined by a speed proportional to both the interaction range and strength \cite{Lieb1972}. 
Suppose two operators $A$ and $B$ are supported on single sites separated by a distance $r$. Then the Lieb-Robinson bound states that, after time $t$, $\left\| \left[A(t), B \right] \right\| \leq c \left\| A \right\| \left\| B \right\| e^{vt - r}$ where $c$ is a constant, $v$ is another constant known as the Lieb-Robinson velocity, and $\| \cdot \|$ represents the operator norm. 
If a system initially in a product state begins evolving under a short-range Hamiltonian, correlations decrease exponentially outside of the causal cone defined by $r = vt$ \cite{Nachtergaele2010,Poulin2010,Nachtergaele2011}. However, in physical systems including polar molecules \cite{Yan2013a,Gorshkov2011,Baranov2012}, Rydberg atoms \cite{Saffman2010,Browaeys2016}, or trapped ions \cite{Islam2013,Porras2004}, the interactions fall off with distance $r$ as a power law $1/r^\alpha$. For these interactions, generalizations of the Lieb-Robinson bound are known, but they may not be tight \cite{Hastings2006,Gong2014,Foss-Feig2015}. In addition, 
for sufficiently long-ranged interactions the causal region may even encompass infinite space at finite time, signaling a breakdown of emergent locality \cite{Eisert2013,Hauke2013,Richerme2014,Metivier2014}. 

These bounds on entanglement have direct implications for quantum information processing. 
The Lieb-Robinson bound, even if time dependence is allowed \cite{Bravyi2006a, Bachmann2012}, limits the speed at which operations can be performed or states created using local Hamiltonians, including states with important applications in quantum  metrology and communication \cite{Gottesman2012,Komar2014,Kessler2014,Bollinger1996,Eldredge2016}. In this paper, we consider the task of using long-range interactions to speed up certain quantum information processes, such as quantum state transfer, GHZ (Greenberger-Horne-Zeilinger) state preparation, and MERA (multiscale entanglement renormalization ansatz) construction. 

State transfer is a process by which an unknown quantum state on one site in a lattice is transferred to another site \cite{Cirac1997,Christandl2004,Bose2003,Bose2007a}. Discussion of possible experimental realizations can be found in Refs.~\cite{DeMoraesNeto2016,Farooq2015,Burgarth2005}, and in Ref.~\cite{Gualdi2008} a case with long-range interactions is considered. Since state transfer establishes perfect correlation between one site at $t = 0$ and another site after the transfer, it is limited by the Lieb-Robinson bound. In this work, we propose a state transfer protocol which makes use of long-range interactions to transfer a state a distance $L$ on a $d$-dimensional lattice in time proportional to $L^0$ $(\alpha < d)$, $\log L$ $(\alpha = d)$, $L^{\alpha - d}$ $(d < \alpha \leq d +1 )$, or $L$ $(\alpha \geq d)$. As an intermediate step of the protocol presented, a GHZ-like state is created, a process also limited by the Lieb-Robinson bound \cite{Bravyi2006a}. 
For polar molecules, Rydberg atoms, or other dipole-dipole interactions in three dimensions, the protocol yields an exponential speed-up in the rate of entanglement generation. 

As we will discuss, one powerful application of fast state transfer using long-ranged interactions would be the realization of a circuit described by a MERA \cite{Vidal2007,Vidal2008,Giovannetti2008}. 
MERAs are particularly useful ways to represent entangled states \cite{Evenbly2009,Pfeifer2009,Swingle2014}, such as the ground states of the toric or Haah codes, topological insulators, and quantum Hall states \cite{Aguado2008,Haah2014,Wen2016,Swingle2016}. 
By performing state transfer and then applying a two-qubit gate between nearest neighbors, we can speed up long-range two-qubit gates, which we use to upper bound the minimal time required to create a MERA state.
Using dipole-dipole interactions in 3D, our protocol constructs the MERA state exponentially faster than using nearest-neighbor interactions. 


\textit{State Transfer.}---Our state transfer protocol first creates a many-body entangled state including the intended starting and final qubits. We do so by applying a controlled $X$ rotation between pairs of qubits $(i,j)$ using a Hamiltonian
\begin{equation}
	H_{ij} = h_{ij} \left( \ket{0} \bra{0}_i \otimes I_j + \ket{1} \bra{1}_i \otimes X_j \right).
	\label{eqn:hamiltonian}
\end{equation}
Here $h_{ij}$ is the interaction strength, which may not be identical for all pairs of qubits. 
In the Supplemental Material, we examine a case where the sign of $h_{ij}$ is variable \footnote{See the Supplemental Material for more details on implementation of the protocol in a dipolar system, which includes References~\cite{Jaksch2000,Saffman2016,Maller2015,Wilk2010,Isenhower2010,DeMille2002,Cirac1995,Wang2015,Huber2011,Merrill2014}.}, but for now we take $h_{ij} > 0$.
$I_j$ and $X_j$ are the identity and Pauli $X$ operator acting on qubit $j$. When the Hamiltonian in Eq.~\eqref{eqn:hamiltonian} is applied for a time $t = \pi/ (2 h_{ij})$, it realizes a controlled-NOT (CNOT) gate between qubits $i$ and $j$ (up to an unimportant phase). In Eq.~\eqref{eqn:hamiltonian}, $i$ is the control qubit for the CNOT while $j$ is the target qubit. When applied to a control qubit in an arbitrary state and a target qubit in the state $\ket{0}$, the CNOT gate results in a two-qubit state encoding the original qubit,
\begin{equation}
	\mathrm{CNOT} \left( a \ket{0} + b \ket{1} \right) \ket{0} = a \ket{00} + b \ket{11}.
\end{equation}
By continuing this process, we can create a many-body entangled state of $N$ qubits $a \ket{0}^{\otimes N} + b \ket{1}^{\otimes N}$ encoding the same state as the initial qubit. The original state can be transferred onto the target qubit by reversing the entangling process and leaving the destination qubit as the final control qubit.
If $H_{ij}$ were a nearest-neighbor Hamiltonian, then this procedure would then allow us to transfer a qubit state a distance $L$ by applying $L$ CNOT operations to construct the many-body state and then $L$ other CNOT operations which are properly time-reversed and spatially mirrored, providing a linear scaling which saturates the Lieb-Robinson bound. 

\begin{figure}[tb]
	\includegraphics[width=8.4cm]{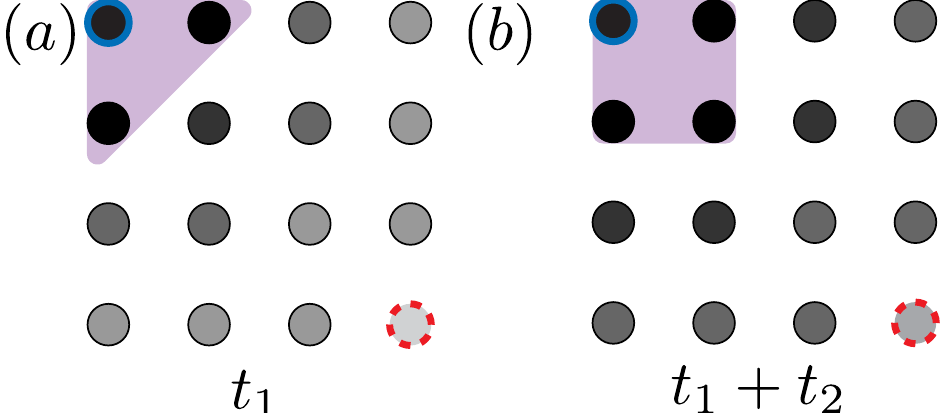}
	\caption{Our state transfer protocol using long-range interactions. We want to move a qubit state from the upper-left site (outlined in solid blue) to the lower-right one (outlined in dashed red).  
	After a time $t_1$ (a), the nearest-neighbor qubits have shifted from target to control (purple region), and continue acting on all other qubits, thereby adding an additional qubit to the set of controls after further time $t_2$, as shown in (b). After $t_2$, each qubit has rotated further (shown by darker shading). The growth continues until the original qubit has effectively performed a CNOT on all qubits in the lattice shown.}
	\label{fig:protocol}
\end{figure}

By using Hamiltonians with long-range interactions, we can achieve a sublinear state transfer time. We suppose that $h_{ij} = 1 / r_{ij}^\alpha$, where $r_{ij}$ is the distance between the qubits $i$ and $j$ \footnote{For $\alpha \leq d$, the thermodynamic limit is not well defined unless the Hamiltonian contains a volume-dependent prefactor proportional to $1/L^{d-\alpha}$ for linear system size $L$ (or $\ln L$ if $\alpha = d$) \cite{Campa2009,Cannas1996}. The inverse of this factor would multiply the required state transfer time. For many physical systems such as polar molecules, this mathematical point will not modify the actually existing interactions over distances of interest, so we do not consider it here.}. 
Our protocol (Fig.~\ref{fig:protocol}) starts by acting on all qubits in the lattice with a single control qubit storing the initial state. 
Once the CNOT operation completes on a qubit, it can be switched from a target to a control and then used to speed up the CNOTs which are still continuing on other qubits.  If a single qubit is targeted by many control qubits, then the CNOT operation on that qubit can be completed faster. (Multiple $H_{ij}$ will mutually commute as long as the sets of target qubits and control qubits are disjoint.) 
If qubit $j$ is targeted by many qubits indexed by $i$, the time required to complete the CNOT becomes
\begin{equation}
	t = \frac{\pi}{2 \sum_{i} h_{ij}} = \frac{\pi}{2 \sum_i r_{ij}^{-\alpha}}.
\end{equation}
(By using dimensionless couplings $h_{ij} = 1/r_{ij}^\alpha$, we are implicitly giving times in units of the inverse nearest-neighbor coupling strength.) In addition to the progressive inclusion of more control qubits, each subsequent qubit has already been rotated by some angle, reducing the remaining time required to complete the operation. Therefore, additional qubits can be added more quickly to the state as it grows.

 As an example, consider beginning with a system of three qubits arranged in a line,
\begin{equation}
	\ket{\psi(t = 0)} = \left(a \ket{0} + b \ket{1} \right) \ket{00}.
\end{equation}
Simultaneously applying $H_{12}$ and $H_{13}$ for a time $t_1 = \pi/2 $, the state becomes
\begin{equation}
	\ket{\psi(t_1)} = a \ket{000} - i  b \ket{11} \left( \cos \frac{ \pi}{2^{\alpha+1}} \ket{0} - i \sin \frac{\pi}{2^{\alpha+1}} \ket{1} \right).
\end{equation}
At this point, the second qubit is made a control, so that the acting Hamiltonians are $H_{13}$ and $H_{23}$. By continuing the evolution under these Hamiltonians for an additional time,
\begin{equation}
	t_2 = \frac{ \frac{\pi}{2} - \frac{\pi}{2 \cdot 2^\alpha} }{ 1 + \frac{1}{2^\alpha}} = \frac{\text{rotation remaining}}{\text{sum of interactions}},
\end{equation} the system will end in the final state
\begin{equation}
	\ket{\psi(t_1 + t_2)} = a \ket{000} - b \ket{111}.
\end{equation}
The entire procedure can be reversed, interchanging the roles of qubits 1 and 3, to transfer the original state,
\begin{equation}
	\ket{\psi\left(2 \left(t_1 + t_2 \right) \right)} = \ket{00} \left( a \ket{0} + b \ket{1} \right).
\end{equation}

\begin{figure}[tb]
	\centering
	\includegraphics[width=8.4cm]{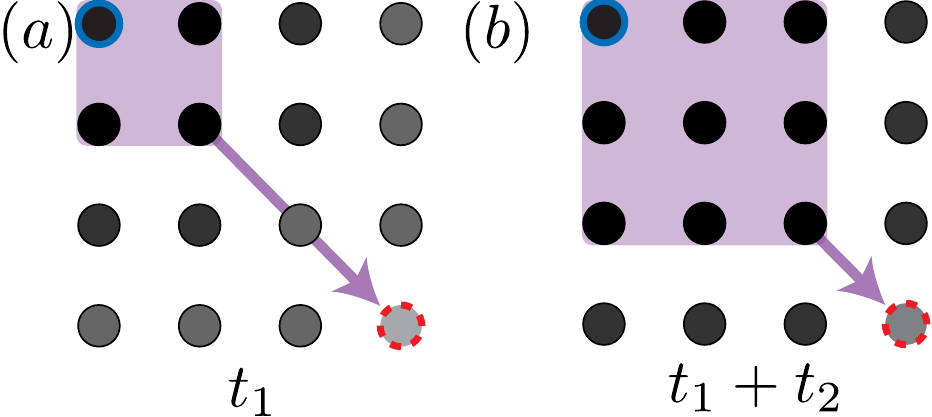}
	\caption{ (a) The suboptimal protocol used for our bounds, with the same color scheme as Fig.~\ref{fig:protocol}. After the $p$th time step, a $(p+1) \times (p+1)$ hypercube of qubits act as controls. The purple arrow represents $H(2,3)$, as it connects a $2 \times 2$ square to a qubit at coordinates $(3,3)$. (b)  After time $t_1 + t_2$ another set of qubits has been converted from targets to controls. The purple arrow now represents $H(3,3)$.}
	\label{fig:badprotocol}
\end{figure}

We now consider the case of many qubits.
First, we specify that we aim to to construct a GHZ state across a hypercube whose diagonal spans a distance $L\sqrt{d}$. The points on either end of the diagonal are the original and destination sites for state transfer (see Fig.~\ref{fig:badprotocol}). Because the state transfer time using the protocol of Fig.~\ref{fig:protocol} is difficult to compute, we use a slightly slower protocol that allows us to easily estimate the transfer time both analytically and numerically. Rather than change a qubit into a control as soon as its evolution completes, we instead halt a qubit's evolution when its rotation finishes. Once we have enough qubits to form a full hypercube of controls, we expand the control set and continue evolution. This scheme is illustrated in Fig.~\ref{fig:badprotocol}, and we expect it to perform similarly (in terms of the scaling of transfer time) to the scheme in Fig~\ref{fig:protocol}. Let $q = 1, 2, \dots, L$ denote each subsequent expansion of the hypercube, so that after time $t = t_1 + t_2 \dots + t_q$ we can form a complete control hypercube of edge length $q$. The times $t_q$ are determined by the condition that each qubit must accumulate a total phase of $\pi/2$,
\begin{equation}
	\sum_{p = 1}^q H(p , q) t_p = \frac{\pi}{2}.
	\label{eqn:timecondition}
\end{equation}
Here $H(p, q)$ is defined to be the summation of all Hamiltonian strengths $h_{ij}$ for which the control $i$ is in the hypercube with corners $\left( 0, 0 ,0 ,\dots \right)$ and $\left( p-1, p-1, p-1, \dots \right)$ and the target $j$ is at the site $\left(q, q, q, \dots \right)$ at the corner of a larger hypercube containing the first, as illustrated in Fig.~\ref{fig:badprotocol}. 
The qubit $j$ is the slowest-evolving qubit on its layer, so its evolution determines the time required to expand the cube in this scheme. 

At this point, we will begin looking for bounds on the times $t_q$. 
Our first bound arises by noting that for all $p$, $t_{p} > t_{p+1}$. This is because, for each $p$, the quantity $H(p,p)$ is strictly larger than $H(p-1,p-1)$ -- the qubit at $(p, p, \dots, p)$ has more qubits acting on it than its counterpart in the previous step. We use $t_p > t_{p+1}$ to rewrite the phase condition on times in  Eq.~\eqref{eqn:timecondition},
\begin{equation}
	\frac{\pi}{2} \geq t_q \sum_{p=1}^q H(p ,q).
	\label{eqn:spigt}
\end{equation}

\begin{figure}[tb]
	\includegraphics[width=8.6cm]{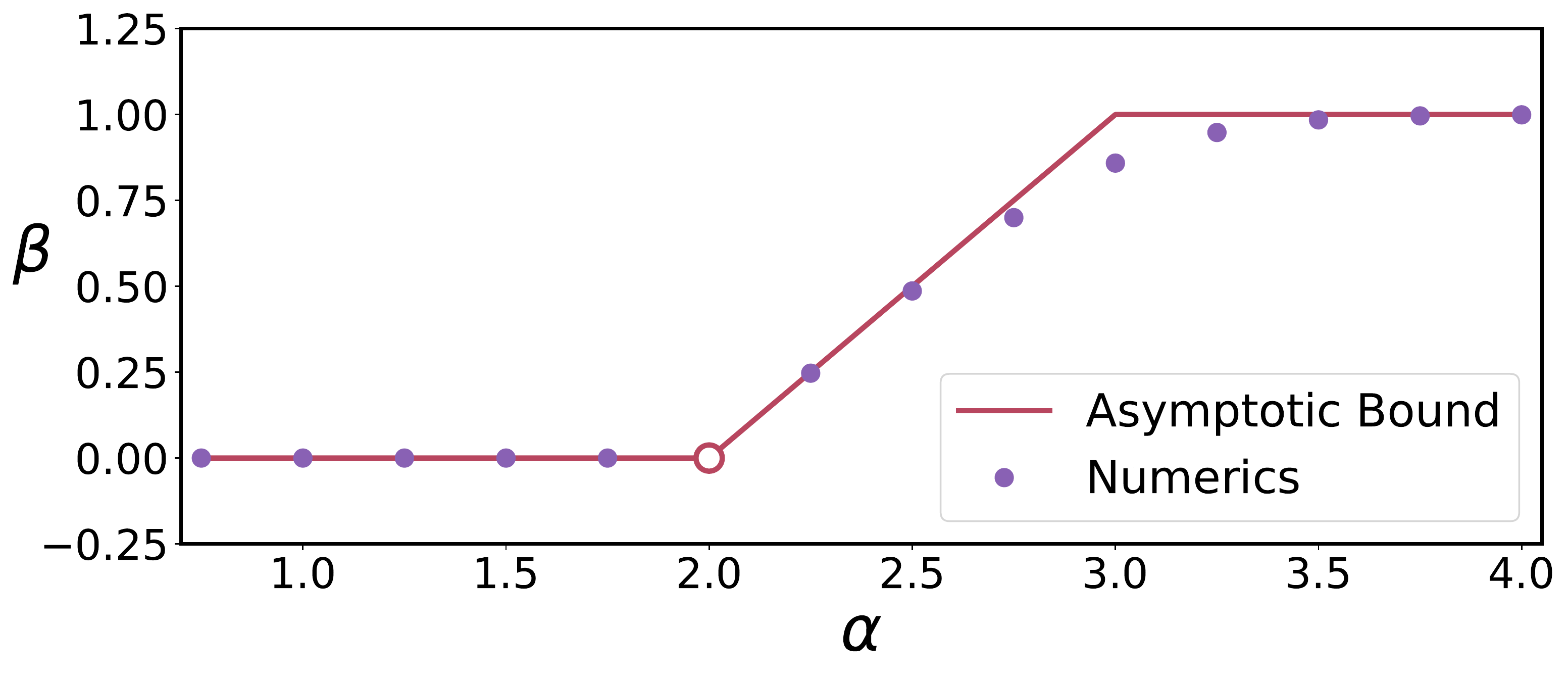}
	\caption{Numerical results of solving Eq.~\eqref{eqn:timecondition} at different $\alpha$ in $d = 2$.  
We calculate $\sum_{q\leq L} t_q$  and fit to $L^\beta$ for $L$ between 900 and 1000; the best-fit exponent is plotted here. 
 The solid line shows the $\beta$ derived from Eq.~\eqref{eqn:tkbound}. At $\alpha = d$ (open circle), the numerics are consistent with the expected logarithmic scaling; the fact that the bound is not saturated at $\alpha = 3$ is due to finite $L$ and should vanish in the $L \to \infty$ limit.}
	\label{fig:scalinggraph}
\end{figure}

We now construct two complementary bounds for $H(p,q)$. In some cases (small $\alpha$), $H(p,q)$ will receive appreciable contributions from the entire hypercube of control qubits. In this case, we can obtain a lower bound by pretending that all control qubits are at the same point a distance $q \sqrt{d}$ away, the maximum possible. However, for large $\alpha$ the interaction is dominated by nearby qubits, whose contributions are independent of $q$. For instance, in $H(q,q)$ there is always one qubit at the nearest vertex of the hypercube whose contribution does not depend on $q$. These two bounds can be combined to yield:
\begin{equation}
	H(p ,q) \geq \max \left( \frac{p^d}{\left(q \sqrt{d} \right)^\alpha}, \frac{\delta_{pq}}{d^{\alpha/2}} \right).
	\label{eqn:sprecountbound}
\end{equation}
After substituting Eq.~\eqref{eqn:sprecountbound} into Eq.~\eqref{eqn:spigt}, the sum can be performed. If we discard all constants depending only on $d$ or $\alpha$, the result is a bound on the scaling of $t_q$,
\begin{align}
	t_q \leq \min \left( q^{\alpha - \left(d + 1\right)}, 1 \right).
	\label{eqn:tkbound}
\end{align}
To obtain the scaling of the entire state transfer process, a sum over $t_q$ is made up to $q = L$. For $\alpha < d$, $t_q$ grows more slowly than $q^{-1}$, so the sum converges to a constant for asymptotic $q$. The convergence signals that a state can be transferred any desired distance in a constant time. For $\alpha = d$, $t_q = q^{-1}$, so the sum scales logarithmically in $L$. For $d < \alpha < d+1$, we obtain a polynomial scaling $L^{\alpha - d}$. Finally, for $\alpha \geq d + 1$, the constant lower bound on $t_q$ dominates, and state transfer takes a time proportional to $L$, just as it does for short-range interacting systems. These scalings are illustrated in Fig.~\ref{fig:scalinggraph} along with the exponents of polynomial fits to the numerical solutions of Eq.~\eqref{eqn:timecondition}. The time cost of our protocol compares very favorably to the direct use of the long-range interaction, which can create a maximally entangled state in time that scales like $L^\alpha$. Note that although Hamiltonians turn on and off throughout our protocol, our Hamiltonians always obey the condition that $\abs{h_{ij}} \leq r_{ij}^{-\alpha}$, meaning that the process as a whole obeys the conditions assumed in previous work on speed limits in long-range interacting systems such as Ref.~\cite{Foss-Feig2015}.


\textit{Constructing a MERA.}---We now demonstrate that our state transfer protocol allows for fast construction of a MERA.  

\begin{figure}[tb]
	\includegraphics[width=8.6cm]{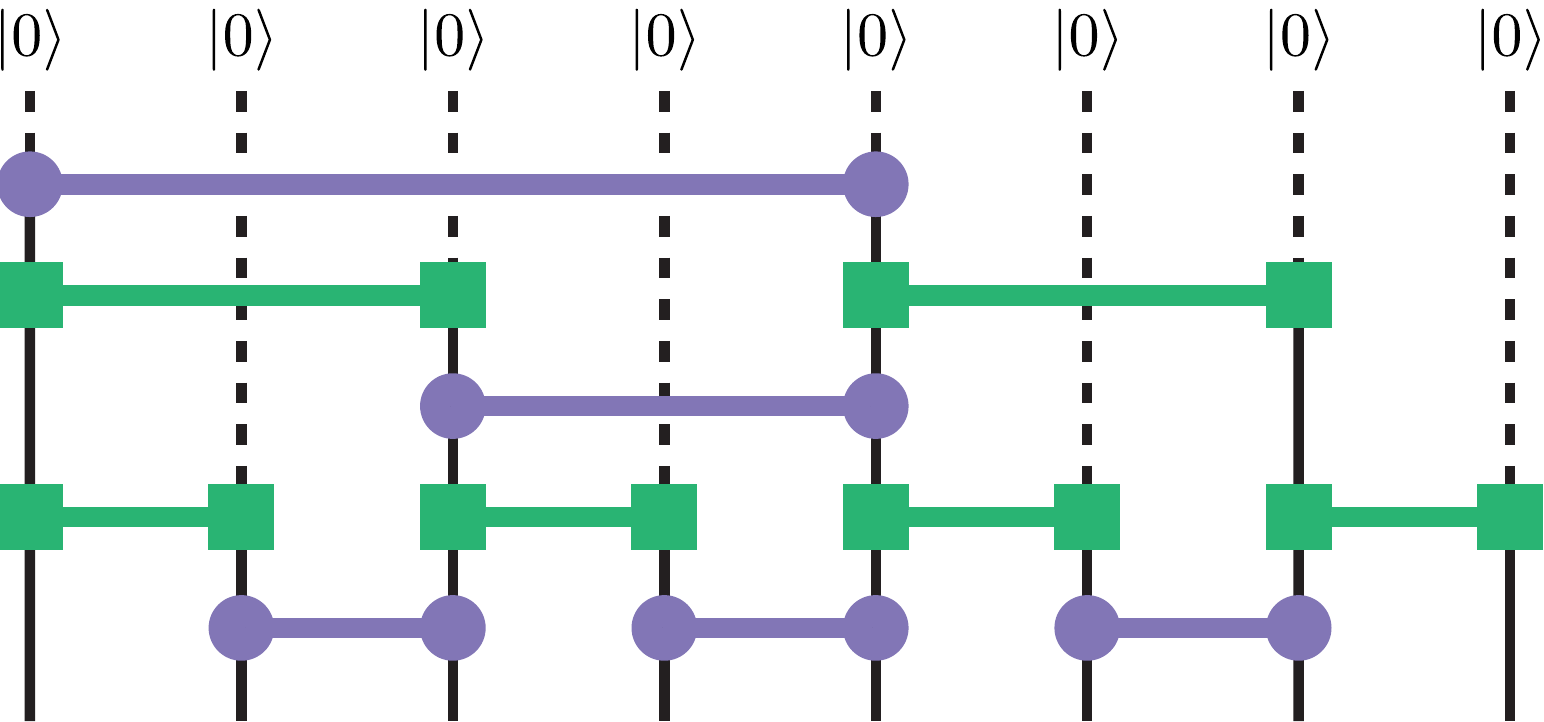}
	\caption{Sketch of a MERA circuit, with the disentanglers (purple, circle-capped) and isometries (green, square-capped). All qubits begin in the state $\ket{0}$, indicated by a dashed line. At each length scale, entanglement is created or removed to create a many-body entangled state from a product state after $\log_2 L$ steps. Although we have drawn a simple 1D binary MERA, our protocol can be extended to higher dimensions and more complicated tensor structures.}
	\label{fig:meracircuit}
\end{figure}

In this context, we will interpret a MERA as a quantum circuit for qubits which acts on successively larger length scales, as shown in Fig.~\ref{fig:meracircuit}, to produce an entangled state from a product state. More general constructions are possible (e.g.\ with qudits). 
Our protocol will also apply to a branching MERA \cite{Evenbly2014a} provided that after a constant number of layers the circuit disentangles a constant fraction of the remaining qubits to $\ket{0}$. This condition ensures that there are always sufficient ``empty'' qubits for our state transfer protocol to scale properly.

A MERA consists of two alternating types of unitary operations and is easiest to understand in reverse (starting at the bottom of the circuit).
The first type of unitary, called a disentangler, removes entanglement at the current length scale. The next operation, an isometry, maps a group of $\phi$ sites into a single site, leaving the other qubits in the state $\ket{0}$. 
These operations can be repeated, except that now all unitary gates need to be performed over a distance $\phi$ times larger than previously.

It is clear that MERA produces a circuit with depth $\log_\phi L$, but this apparent logarithmic scaling masks an actual time cost due to the continuously increasing length scale. However, we can replace a long-range two-qubit unitary with state transfer followed by a short-range unitary. This framework allows us to ignore any details of the two-qubit unitary and simply use state transfer as a primitive. The structure of a MERA circuit guarantees that the $\ket{0}$ states required to perform state transfer will be present between any two qubits when we need to perform a unitary on them. 

Suppose that $t_\tau$ is the maximum time required to perform a two-qubit gate across a distance $\ell_\tau$ at the $\tau$th step of the MERA circuit. We can perform all the MERA operations at a given step in parallel, so a single layer of the MERA simply requires time $2 t_\tau$ for the disentanglers and then isometries.
The time to perform the entire MERA circuit will then be bounded (up to a constant factor) by
\begin{equation}
	t_\mathrm{MERA} \lesssim \sum_{\tau=0}^{S-1} t_\tau.
	\label{eqn:layerscaling}
\end{equation}
Here $S = \log_\phi L$. Our state transfer procedure allows for $t_\tau = 2 t_\mathrm{transfer}$. The time required to perform the final two-qubit gate does not affect the scaling and so is omitted. We can then bound $t_\mathrm{transfer}$ by considering the length scale at each step, $\ell_\tau = \phi^\tau$. If $\alpha = d$, $t_\mathrm{transfer}$ scales as $\log_\phi \ell_\tau$ (as in our state transfer bound but with a constant multiple changing the base of the logarithm), and $t_\mathrm{MERA}$ will be bounded by  $\sim \left( \log_\phi L \right)^2$ by considering the largest term in Eq.~\eqref{eqn:layerscaling} multiplied by the number of terms. For $\alpha \neq d$, $t_\mathrm{transfer}$ scales polynomially in $\ell_\tau$ with exponent $\beta$, 
\begin{equation}
	t_\mathrm{MERA}  \lesssim \sum_{\tau=0}^{S-1} \ell_\tau^{\beta}.
\end{equation}
For $\alpha < d$, $\beta = 0$ and the sum is proportional to $\log_\phi L$. For $\alpha > d$, $\beta = \max \left( \alpha - d, 1 \right)$. We use $\ell_\tau = \phi^\tau$ and carry out the geometric sum to obtain 
\begin{align}
	t_\mathrm{MERA} &\lesssim \left( \phi^{\beta} \right)^S =  L^{\beta} .
\end{align}
Thus we have
\begin{equation}
	t_\mathrm{MERA} \lesssim \begin{cases} \log_\phi L & \alpha < d \\ \log_\phi^2 L & \alpha = d \\ L^{\alpha - d} & d < \alpha \leq d + 1 \\ L & \alpha > d + 1. \end{cases}
\end{equation}

\textit{Outlook.}---We have demonstrated fast state transfer and MERA construction protocols using long-range interactions.
Our protocol’s exponential speedup for $\alpha = d$ nearly saturates the bound in \cite{Hastings2006}, which gives a logarithmic lightcone for $\alpha > d$.
However, we have not shown that our method is the fastest state transfer protocol possible. Such a result would require demonstrating a general Lieb-Robinson-type bound which we would then saturate. 
Instead, our protocol limits future Lieb-Robinson bounds for long-range systems. The state transfer protocol we have presented establishes that no finite causal region is possible for $\alpha < d$, since a constant amount of time suffices to establish any desired correlation at arbitrary distances. In previous work, causal regions were seen in systems with $d/2 < \alpha \leq d$ as long as the initial state was not entangled \cite{Eisert2013}. Like our work, Ref.~\cite{Eisert2013} also uses multiple qubits with long-range interactions to reduce state transfer time. We have shown that such causal regions do not persist in general, although it is possible that this violation requires the use of time-dependent Hamiltonians as opposed to the time-independent Hamiltonians in Ref.~\cite{Eisert2013}. 

For the intermediate value $d < \alpha < d + 1$, our protocol shows that no linear light cone can be drawn, although a polynomial bound may be possible. 
These results should be compared to Ref.~\cite{Foss-Feig2015}, which established a polynomial light cone only for $\alpha > 2D$ that becomes linear only in the limit of $\alpha \to \infty$.
Our protocol's linear scaling when $\alpha \geq d + 1$ suggests that the tightest possible Lieb-Robinson bound may also possess a critical $\alpha$ with a similar property. Resolving this question could reveal important facts about the nature of correlations in long-range interacting systems.

It is our hope that this protocol, or a minor variation thereof, could soon be realized experimentally. Such a realization could offer significant technological advantages in, for instance, entanglement-enhanced metrology. In the Supplemental Material, we show how dipole-dipole interactions in three dimensions can be used to implement a variant of our protocol with a focus on Rydberg atoms \cite{Note1}. Using this protocol, qubits can be entangled exponentially faster than using short-range interactions. In the future, we hope to reduce the local control required to achieve sublinear scaling. 

We thank G. Evenbly and N. Yao for discussions.
This work was supported by the AFOSR, ARO MURI, ARL CDQI, NSF QIS, ARO, and NSF PFC at JQI. Z. E. is supported in part by the ARCS Foundation. 
\bibliography{./stlib.bib}{}

\onecolumngrid
\appendix 
\section{Application to Dipole-Dipole Interactions}
In this Supplement, we show that it is possible to realize a protocol similar to that in the main text by using Rydberg atoms. Rydberg atoms can be made to interact with a dipole-dipole interaction that has distance dependence $1/r^3$. This suggests that, using our protocol, we could produce a cube of side length $L$ in a GHZ state in time proportional to $\log L$. We will demonstrate that a realistic physical interaction can yield this result. Many details on Rydberg atoms and their applications in quantum information can be found in Refs.~\cite{Saffman2010,Jaksch2000,Saffman2016}, and experimental demonstrations can be found in Refs.~\cite{Maller2015,Wilk2010,Isenhower2010,Browaeys2016}. Our analysis is focused on Rydberg atoms, but much of it should extend to other dipolar systems, such as polar molecules, with appropriate modification of implementation details \cite{Gorshkov2011,DeMille2002,Yan2013a,Baranov2012}.  

We select as qubit states the ground state and a highly excited state of a Rydberg atom under a weak electric field, yielding a purely diagonal atomic interaction \cite{Saffman2010}. The Hamiltonian of a system of such atoms can be written as:
\begin{equation}
	H_\mathrm{int} = \sum_{i \neq j } H_{ij} = \sum_{i \neq j}  \frac{1 - 3 \cos^2 \theta_{ij}}{r_{ij}^3} Z_i Z_j \equiv \sum_{i \neq j} V_{ij} Z_i Z_j.
	\label{eqn:hint}
\end{equation}
Here, $r_{ij}$ is the distance between atoms $i$ and $j$, while $\theta_{ij}$ is the angle between the electric field and the vector separating the two atoms. We have ignored local terms like $Z_i$ and $Z_j$, which can be removed by applying local rotations. By applying local rotations, this $ZZ$ Hamiltonian can be used to realize CNOT interactions, regardless of whether the overall sign is positive or negative. This is done by applying local rotations to produce a controlled-phase gate and applying Hadamard operations on the target before and after the evolution to yield a controlled-NOT gate \cite{Cirac1995}. We assume that, while local control fields may be time-dependent, the two-body interaction in Eq.~\eqref{eqn:hint} is active throughout the entire state transfer process. The individual addressing required to perform these local operations was demonstrated in a 3D optical lattice in Ref.~\cite{Wang2015}. The roughly 5 $\mu \mathrm{m}$ lattice spacing in that work is also an appropriate spacing for the Rydberg interactions we intend to use in our protocol, as it helps to prevent the dipole-dipole interactions from becoming comparable to the energy level spacing.

To apply the protocol in the main text, qubits must be separated into controls and targets. Such separation can be performed using an echoing procedure: first, qubits evolve under $H_\mathrm{int}$ and then under $-H_\mathrm{int}$ for an equal amount of time. However, halfway through the second evolution, a $\pi$-pulse ($X$ gate) is applied to either all target qubits or all control qubits. This has the effect of swapping $Z$ for $-Z$. All interactions between controls and controls, or targets and targets, will remain unchanged, but any control-target interactions will be inverted. Thus, during the $-H_\mathrm{int}$ time, control-target interactions experience no net evolution, while any control-control or target-target pair evolution due to $+H_\mathrm{int}$ is undone. The $-H_\mathrm{int}$ evolution time is equal to the initial entangling $+H_\mathrm{int}$ time, so the echoing procedure does not change the scaling with $L$. Even if the negative interaction is not of the same magnitude as the original, we can still accomplish the echoing by adjusting the timescales, and the scaling with $L$ will still not be changed. 


To change the sign of the dipole-dipole interaction, realizing $-H_\mathrm{int}$, we can encode the computational states into the fine structure of a Rydberg atom. For specificity, we consider the case of $\mathrm{Rb}^{87}$ with a weak applied electric field. Ignoring the hyperfine structure, we encode the state $\ket{0}$ in a superposition of $\ket{L = 0, J = 1/2, m_J = 1/2}$ and $\ket{L = 1, J =3/2, m_J = 3/2}$ created by applying a microwave dressing field, with most of the amplitude being stored in the latter state. The state $\ket{1}$ is then encoded in $\ket{L = 1, J = 1/2, m_J = 1/2}$. All three states have the same principal quantum number.
Details can be found in an analogous scheme for polar molecules presented in entry No. 5 of Table II and Fig 3(d) of Ref.~\cite{Gorshkov2011}. Note that here we are also dropping local $Z$ terms which can be canceled by a local rotation. 
We have calculated dipole matrix elements for $\mathrm{Rb}^{87}$ across a wide range of principal quantum numbers that confirm this scheme remains viable in the Rydberg setting. We also assume that, in addition to changing the overall sign of the interaction, we are able to place qubits in non-interacting electronic ground states to avoid any unwanted interactions or decay from excited states. 

If a volume of control qubits exists, this volume will convert a qubit $j$ into a control after time $\pi / 2 V_j$, where $V_j$ is the sum over all interaction constants $V_{ij}$ for control qubits $i$. Suppose that enough qubits have been added that the sum of point-to-point interactions is well-approximated by an integral, which is a good approximation in the relevant asymptotic regime. The total interaction on a qubit $j$ in this case can be written as
\begin{equation}
	V_j  = \int_C   V_{ij} \dd{C}.
	\label{eqn:vint}
\end{equation}
Here, $C$ is the volume of control qubits. This quantity has the useful property of scale invariance. If all lengths change by a factor $\lambda$, then $H_\mathrm{int}$ changes by the factor $\lambda^{-3}$ due to its distance dependence. However, the region of integration expands by $\lambda^3$, so the final quantity remains unchanged.

We consider expanding a cube of controls, increasing the side length $\ell$ by a constant factor $\lambda$. After this procedure, we obtain a new cube of side length $\lambda \ell$. Qubits outside of the larger cube have no operations performed on them. Once this expansion has been performed, we expand the cube again. Due to scale invariance, the same operation can be performed in identical time. This means that after $n$ expansion steps, the side length will be $\lambda^n \ell$. Therefore, we can construct a cube of side length $L$ in a time proportional to $\log_\lambda \left( L/\ell \right)$ as indicated in the main text. The scaling properties of the integral in Eq.~\eqref{eqn:vint} can be used in cases where $\alpha \neq d$ as well. Equation~\eqref{eqn:vint} implies that the time required to construct a cube of side length $L$ will be:
\begin{equation}
	t_\mathrm{GHZ} \sim \sum_{i = 1}^{log_\lambda \left( L / \ell \right)} \lambda^{n ( \alpha - d )}.
\end{equation}
For $\alpha < d$, this saturates to a number independent of $L$, and for $\alpha > d$, it implies that $t_\mathrm{GHZ} \sim L^{\alpha - d}$. Note that for $\alpha > d +1$, a protocol of successive dilations of the cube fails to provide optimal scaling.

\begin{figure}[tb]
	\includegraphics[width=.7\textwidth]{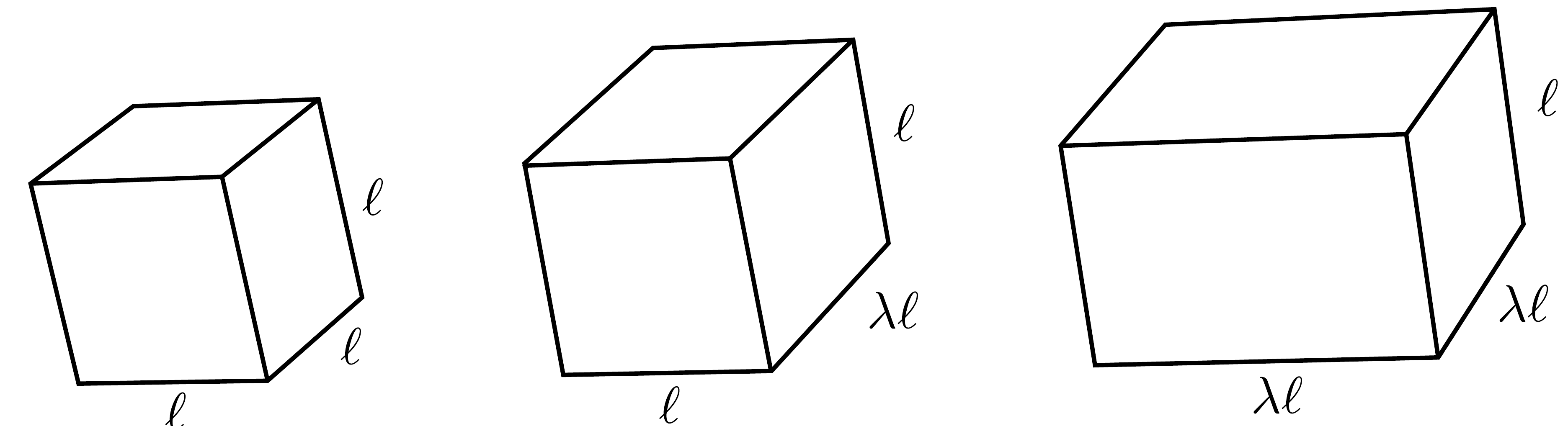}
	\caption{Successive transformations of the control cube. A cube of side length $\ell$ is expanded first in one direction, then the next. After the final step (not shown), the result will be a cube of side length $\lambda \ell$.}
	\label{fig:cubes}
\end{figure}

All that remains to be shown is that the size of the cube can be increased by a constant factor in finite time. This is not guaranteed because the dipole-dipole interaction changes sign as a function of $\theta_{ij}$, causing $V_j$ to be zero for qubits at some points. If we could only act with the control cube during the expansion time, we would not be able to perform the expansion as outlined above. However, we can use a slightly more complicated scheme in which some intermediate qubits are used. Rather than expand the entire cube at once, we expand the cube outward in the positive $x$-, $y$-, and $z$-directions successively, each time expanding only to qubits which lie on lines perpendicular to the expanding face of the rectangular prism, as illustrated in Fig.~\ref{fig:cubes}. This works because the interaction can be shown to decrease monotonically (in absolute value) along Cartesian directions, as we prove below. Since at long distances we know that the interaction decays to zero and has the same sign for all target qubits, the monotonicity establishes that there is no zero crossing. As there is no zero crossing, there will be a finite time that suffices to complete the expansion. The logarithmic scaling follows.

\section{Proof of Interaction Monotonicity}

\begin{figure}[tb]
	\includegraphics[width=.4\textwidth]{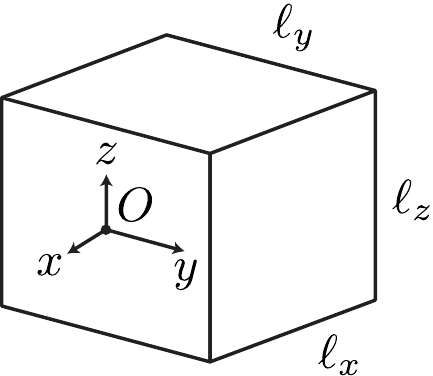}
	\caption{Illustration of the coordinate system used in this section.}
	\label{fig:monotonic}
\end{figure}
We will now prove that the interaction between a cube of controls and a target qubit decreases monotonically in Cartesian directions. Suppose we begin with a rectangular prism located in the $y-z$ plane with dimensions $\ell_x \times \ell_y \times \ell_z$ and the origin in the center of one face (see Fig.~\ref{fig:monotonic} for an illustration). A qubit at point $(x,y,z)$ then has the interaction integral
\begin{equation}
	V = \int_x^{x + \ell_x} \int_{-\ell_y/2 + y}^{\ell_y/2+y} \int_{-\ell_z/2+z}^{\ell_z/2 + z} \frac{x'^2 + y'^2 - 2z'^2}{\left(x'^2 + y'^2 + z'^2\right)^{5/2}} \dd{x'} \dd{y'} \dd{z'}.
	\label{eqn:monointegral}
\end{equation}
The integrand in Eq.~\eqref{eqn:monointegral} is simply the dipole interaction written in Cartesian coordinates. We choose $y$ and $z$ to fall in $(-\ell_y/2, \ell_y/2)$ and $(-\ell_z/2,\ell_z/2)$ respectively to ensure that their projection to the $y-z$ plane lies on the face of the prism. We consider only positive values of $y$ and $z$ without loss of generality. The derivative of $V$ with respect to $x$ can be expressed analytically as
\begin{align}\label{eqn:dx1}
	\partial_x V &= D\left( - \frac{\ell_y}{2} + y, - \frac{\ell_z}{2} + z \right) + D\left( \frac{\ell_y}{2} + y,  \frac{\ell_z}{2} + z \right) \\ \nonumber 
	&- \left[D\left(- \frac{\ell_y}{2} + y,  \frac{\ell_z}{2} + z \right) + D\left(\frac{\ell_y}{2} + y, - \frac{\ell_z}{2} + z \right) \right], 
\end{align}
\begin{align}
	D(a,b) &= a b \left( \frac{1}{ \left( \left(x + \ell_x \right)^2 + c^2 \right) \sqrt{ \left( x + \ell_x \right)^2 + a^2 + c^2 }} - \frac{1}{ \left( x^2 + c^2 \right) \sqrt{ \left( x^2 + a^2 + c^2\right) }}  \right).
\end{align}
For $D(a,b)$, the sign is always determined by the prefactor because the factor in parentheses is strictly negative. Using the fact that $y$ and $z$ must be less than $\ell_y/2$ and $\ell_z/2$ respectively, we can assign a negative sign to the first two $D$ to appear in Eq.~\eqref{eqn:dx1} and a positive sign to the second two. Therefore, we find that $\partial_x V$ is always negative in this region, establishing the monotonicity for expansion along one face in the $x$-direction. This proof also holds for the $y$-direction immediately from symmetry. For the $z$-direction, a similar argument holds but with a more complicated parenthetical term in $D(a,b)$.

\section{Effects of Decoherence}
In the next two sections, we will consider the influence of experimental imperfections in qubits and gate operations and examine the implication for our protocol’s scalability. 
 First, we will consider the influence of decoherence, for instance, due to spontaneous emission out of the Rydberg excited states. The fragile nature of the GHZ state means that a single emission can cause our protocol to fail. We assume that individual qubits fail (spontaneously emit) at a rate $\gamma$. This analysis should extend to any similar failure mechanism that occurs at a constant rate. 
If each expansion step (dilating the cube by $\lambda$) takes time $\delta t$, then we can consider whether, in the $i$th timestep, any of the $\lambda^{3 i}$ qubits currently involved emit. If so, we label the step a success. The protocol succeeds if all of its individual steps succeed. The probability that no spontaneous emissions occur at any of $N_t$ time steps and that the protocol succeeds is
\begin{equation}
	P(\mathrm{success}) = \prod_{i=1}^{i = N_t} P( \text{success at step $i$} ) =  e^{- \gamma \delta t \sum_i \lambda^{3i}}.
	\label{eqn:success}
\end{equation}
If we demand that the protocol successfully entangle $N$ qubits with a probability $P > \epsilon$, then Eq.~\eqref{eqn:success} becomes
\begin{equation}
	\sum_{i=1}^{\log_\lambda N^{1/3}} \lambda^{3i} = \frac{ \lambda^3 \left( N - 1 \right)}{\lambda^3 - 1} <  \frac{\ln \frac{1}{\epsilon}}{\gamma \delta t}.	
	\label{eqn:gatecount}
\end{equation}
This suggests a limit on the number of qubits which can be entangled with a system of decohering qubits, which we write as
\begin{equation}
	N_\mathrm{lr} < 1 + \frac{\ln \frac{1}{\epsilon}}{\gamma \delta t} \frac{\lambda^3 -1 }{\lambda^3}.
	\label{eqn:nlr}
\end{equation}

Here $N_\mathrm{lr}$ refers to the number of qubits that can be entangled using our long-range interacting protocol. Note that if $\epsilon$ and $\lambda$ are taken to be of order 1, Eq.~\eqref{eqn:nlr} simply implies that $N_\mathrm{lr} \gamma \delta_t \lesssim 1$, which is unsurprising since our largest entangled state decays in a time $1 / N_\mathrm{lr} \gamma$.
We can also consider what this limit looks like in the case of a protocol which uses nearest-neighbor interactions and, at each step, increases the cube's side length by one. In this case, the $i$th timestep has $i^3$ qubits entangled, and there are $N^{1/3}$ such steps. A similar argument to the above leads us to calculate
\begin{equation}
	\sum_i^{N^{1/3}} i^3 = \frac{1}{4} \left[ N^{4/3} + 2N + N^{2/3} \right] < \frac{\ln \frac{1}{\epsilon}}{\gamma \delta t}.
	\label{eqn:nnn}
\end{equation}
If we assume we're interested in cases where $N$ is somewhat large \textit{a priori}, then we write the following loose bound by dropping strictly positive terms:
\begin{equation}
	N_\mathrm{nn}  <  \left( \frac{4 \ln \frac{1}{\epsilon}}{\gamma \delta t} \right)^{3/4}.
\end{equation}
Here the exponent $3/4$ arises because we summed over $N^{1/3}$ terms like $i^3$, yielding $N^{4/3}$ and then inverted that. Suppose we take $\lambda = 2$, in which case the first step of each protocol is the same and we can equate the two $\delta t$. Then the ratio of the two threshholds is
\begin{equation}
	\frac{N_\mathrm{lr}}{N_\mathrm{nn}} =  \frac{7}{16 \sqrt{2}} \left( \frac{\ln \frac{1}{\epsilon}}{ \gamma \delta t} \right)^{1/4}.
\end{equation}

To evaluate this figure of merit, we can look at the original proposal for interaction-based Rydberg gates, which suggests a two-qubit gate timescale of less than a nanosecond \cite{Jaksch2000}. Our protocol also requires several one qubit gates in each step, which can also be accomplished on nanosecond timescales using pulsed lasers \cite{Huber2011}. Estimating $\delta t \sim 5 \ \mathrm{ns}$, demanding a success probability of $1/2$, and taking the $\mathrm{Rb}^{87}$ 100s state lifetime of $340 \ \mu\mathrm{s}$ at a temperature of 300K \cite{Saffman2010}, we find that $N_\mathrm{lr} / N_\mathrm{nn} \approx 4.5$, meaning that a long-range protocol can achieve a maximally entangled state containing nearly 4.5 times as many qubits as one constructed by nearest-neighbor interactions. This figure rises to 4.9 if we solve Eq.~\eqref{eqn:nnn} directly rather than using the bound. $N_\mathrm{lr}$ is about $4 \times 10^4$, suggesting a lifetime for the GHZ state of roughly $8 \ \mathrm{ns}$. Using $\delta t$ and $N_t = \log_\lambda N_\mathrm{lr}^{1/3}$, we find that constructing such a state would require a total time of about $25 \ \mathrm{ns}$.

To estimate the probability of performing state transfer instead of constructing the GHZ state, one must simply replace $\epsilon$ with $\sqrt{\epsilon}$ in the above analysis, as a state transfer success is effectively just two successful iterations of the GHZ construction. After state transfer is performed, we can ask whether it survives long enough to be read out or transferred into a non-interacting level. Since the single-atom lifetime of the Rydberg state is 340 $\mu$s, this should not be an issue as the time required to complete the transfer is on the order of tens of nanoseconds. Once transfer or GHZ creation is complete, the electric field can be turned off to remove the dipole-dipole interaction in Eq.~\eqref{eqn:hint}. 

\section{Effects of Imperfect Single-qubit Gates}
In addition to free evolution under the long-range interaction Hamiltonian [Eq.~\eqref{eqn:hint}], our protocol requires a number of single-qubit gates to be performed. These can be Hadamard gates which produce the CNOT operation out of our $ZZ$ interaction or the echoing pulses. In any case, a failure of the single-qubit gate can pose a serious problem to the protocol. Suppose we perform $N_s$ single-qubit gates which succeed with a probability $P$. Then, as in the previous section, we demand that the gate sequence succeed with probability $\epsilon$, obtaining
\begin{equation}
	P^{N_s} > \epsilon \implies P > e^{(\ln \epsilon)/N_s}.
\end{equation}

The number of single qubit gates which must be targeted on a qubit in a timestep varies depending on that qubit's role during the step, but let us suppose that on average there are $c$ gates per qubit performed on each of $N_t$ timesteps. We can count the number of qubits involved in each timestep just as we did in Eq.~\eqref{eqn:gatecount} to obtain a criterion for success:
\begin{equation}
	P > e^{(\ln \epsilon)/(c \lambda^3 (N - 1)/(\lambda^3 - 1))}.
	\label{eqn:oneqg}
\end{equation}
Theoretical work on composite pulse sequences for atomic qubits suggests achievable fidelities of $1 - 10^{-4}$ \cite{Merrill2014}. If we assume $c = 4$ as an estimate, $\epsilon = 1/2$, and $\lambda = 2$ as in the last section, Eq.~\eqref{eqn:oneqg} suggests that roughly 1500 qubits could be entangled with such gates using our protocol. This is a reduction of several orders of magnitude from the previous section which considered no single-qubit fidelity issues, a limitation which highlights the fact that a version of the protocol requiring less single-qubit control could perhaps entangle dramatically more qubits.
\end{document}